 \documentclass[review]{elsarticle}

\usepackage{subfig}
\usepackage{lineno}
\usepackage{multirow}
\usepackage{makecell}
\usepackage{float}

\usepackage{amssymb}

\usepackage{lineno}

\journal{ Nucl. Instrum. Meth. A}
\begin{document}

\begin{frontmatter}



\title{\boldmath Batch test of MRPC3b for CBM-TOF/STAR-eTOF}


\author[address1,address2]{K.Wang\fnref{1}}

\author[address1,address2]{J.Zhou\fnref{1}}

\author[address1,address2]{X.Wang\fnref{1}}

\author[address1,address2]{X.Li}

\author[address1,address2]{D.Hu\corref{correspondingauthor}}
\ead{hurongr@ustc.edu.cn}

\author[address1,address2]{Y.Sun\corref{correspondingauthor}}
\ead{sunday@ustc.edu.cn}
\fntext[1]{These authors contributed equally to this work.}

\address[address1]{State Key Laboratory of Particle Detection and Electronics, University of Science and Technology of China, 96 Jinzhai Road, Hefei 230026, China}
\address[address2]{Department of Modern Physics, University of Science and Technology of China (USTC), 96 Jinzhai Road, Hefei 230026, China}
\cortext[correspondingauthor]{Corresponding author}

\begin{abstract}
The Compressed Baryonic Matter (CBM) experiment is one of the major scientific spectrometers of the future Facility for Antiproton and Ion Research (FAIR) in Darmstadt. As one of the core sub-systems in CBM experiment for charged hadron identification, the Time-of-Flight (TOF) system is required to have a time resolution better than 80 ps. According to the final state particle flux distribution, the CBM-TOF will be constructed with several types of Multi-gap Resistive Plate Chambers (MRPC). In the outer region of the TOF wall where the particle fluxes are around 1 $\rm kHz/cm^2$, MRPCs with ultra-thin float glass electrodes are considered as a cost effective solution. MRPC3b prototypes have been developed and tested with excellent performance which could meet all the requirements. Before the construction of CBM-TOF, approximately 80 MRPC3bs are assembled for the STAR endcap TOF (STAR-eTOF) upgrade at RHIC as part of the FAIR Phase-0 programs for CBM-TOF which provides a valuable opportunity for detector stability test under high flux environments. This paper will introduce the batch test of the MRPC3bs for STAR-eTOF upgrade. Time resolution of better than 70 ps and efficiency of around 95\% are achieved. Notably, during the batch test, it has been observed that the noise rates of the two edge strips in each counter are significantly higher than those of the middle strips. Simulations with Computer Simulation Technology (CST) Studio Suite are carried out and several kinds of MRPC prototypes are designed and tested accordingly.  Based on the simulation and test results, the design of the MRPC3b has been further optimized, resulting in a significant suppression of noise rates in the edge strips.
\end{abstract}




\end{frontmatter}


\section{Introduction} \label{sec:intro}
\begin{figure*}[htbp]
	\centering 
	
	\subfloat[]{\includegraphics[width=0.45\textwidth]{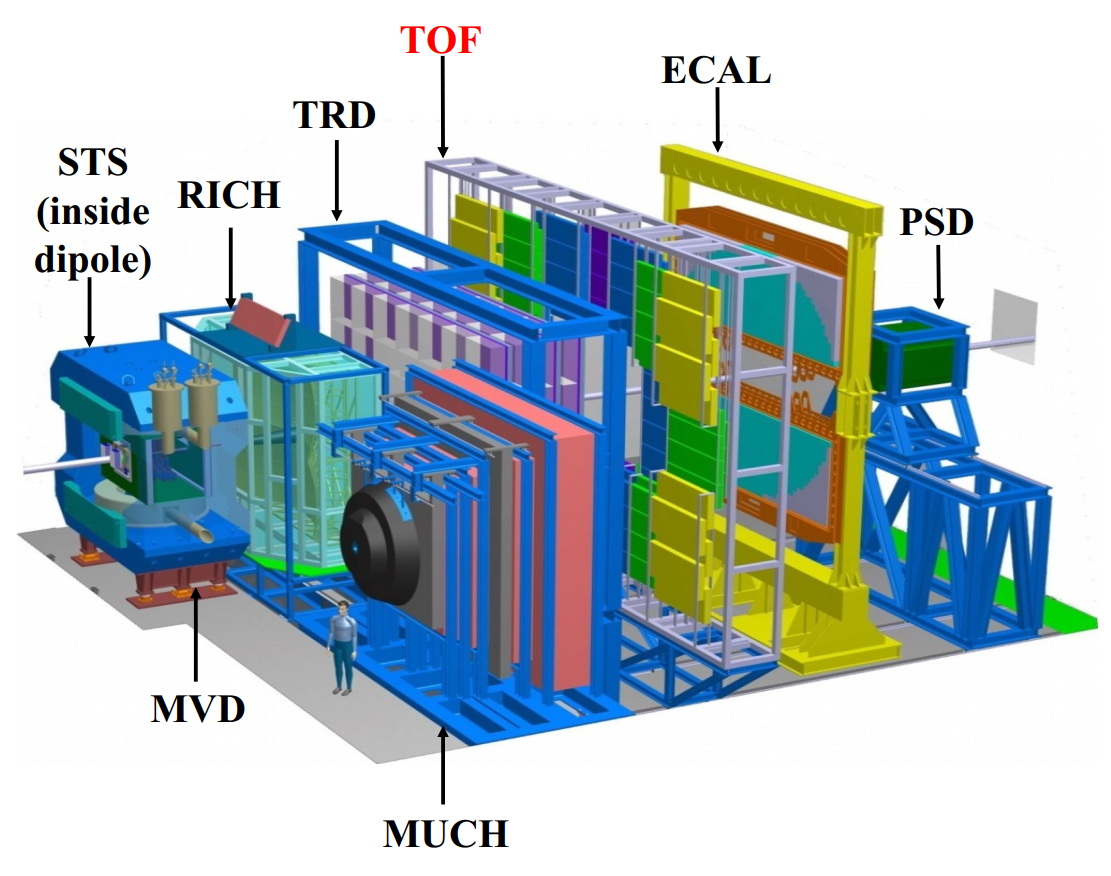}} \quad
	\subfloat[]{\includegraphics[width=0.48\textwidth]{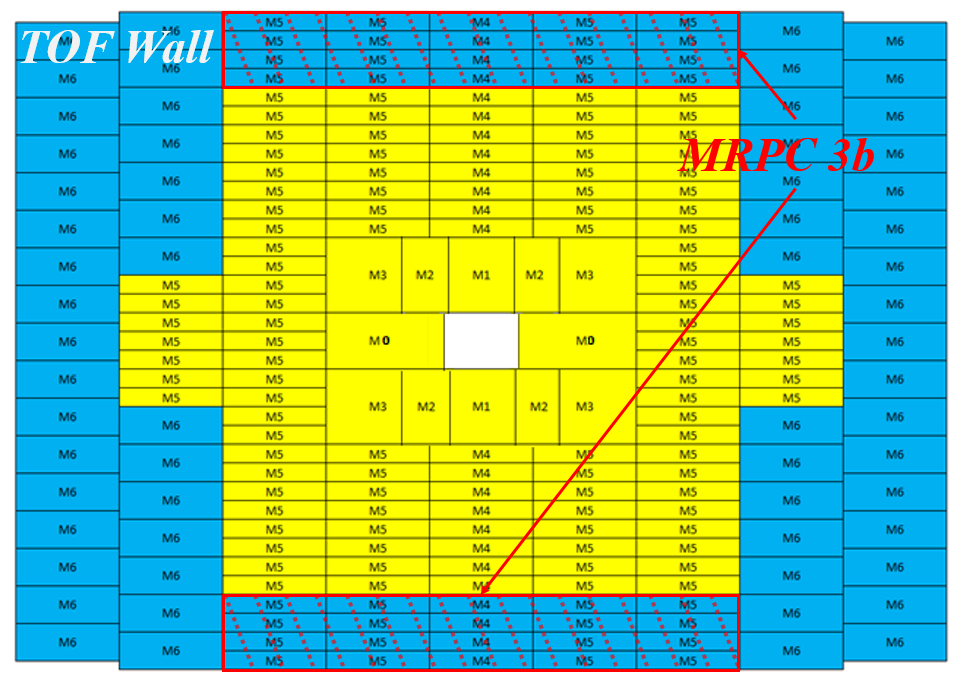}}
	\caption{ (a) The conceptual design of the CBM spectrometer\cite{CBMweb}. (b) The sketch of CBM-TOF wall.}
	\label{fig1}
\end{figure*}

The Compressed Baryonic Matter (CBM) experiment is a future fixed target heavy ion experiment located at the Facility for Antiproton and Ion Research (FAIR) in Darmstadt. The main Physics goal of CBM experiment is to map the phase diagram of strongly interacting matter at high baryon densities through the study of heavy ion collisions in the beam energy range from 2 AGeV to 11 AGeV  \cite{CBM1,CBM}. Figure \ref{fig1}(a) shows the conceptual design of the CBM spectrometer. A 120 $\rm m^2$ Time-of-Flight (TOF) system located at 10 m downstream from the target will provide charged hadron (protons, kaons, pions) identification up to a particle momentum of about 4 ${\rm GeV/}c$ \cite{deppner2019cbm}. The Multi-gap Resistive Plate Chamber (MRPC)\cite{multigap_RPC}, as a gaseous detector with excellent timing performance and low cost, is selected as the basic detector component of the CBM-TOF. According to the flux rates, from tens $\rm kHz/cm^2$ to around one $\rm kHz/cm^2$, the CBM-TOF is divided into several sub-regions, as sketchily shown in Figure \ref{fig1}(b) \cite{TOFTDR}. 
In the outer region marked in blue, which covers over half of the total TOF wall, the estimated particle flux is around 1 $\rm kHz/cm^2$. MRPCs made of ultra-thin float glass electrodes (referred to as MRPC3b and MRPC4 in the CBM TOF TDR) will be employed in this region with appropriate rate capability at economical cost. The MRPC3b prototype has been developed and tested with excellent performance \cite{mrpc3b,mrpc3bbeam}. 

In order to further test the long-term stability under high flux environment, the CBM-TOF modules were installed for STAR endcap TOF (STAR-eTOF)  upgrade as part of the FAIR Phase-0 program at the end of 2018 \cite{toffairphase0}, among which 81 MRPC3bs were constructed and tested by USTC.

This paper presents the design of the MRPC3b counter and details the batch test conducted on the 81 MRPC3bs for the STAR-eTOF upgrade, including the cosmic batch test platform and batch test results. Besides, an issue was observed during the batch test that the noise rates of the two edge strips were much higher than the middle ones. Simulations based on the Printed Circuit Board (PCB) Studio of the Computer Simulation Technology (CST) Studio Suite were carried out, leading to the design and testing of multiple MRPC prototypes. Through these research and development efforts, the optimized MRPC3b structure demonstrates a notable improvement in edge strip noise rates. This optimized design will be implemented in the final construction of the CBM-TOF.

\section{MRPC3b counter} \label{MRPC3b}

\begin{figure}[htbp]
	\centering 
	
	\subfloat[]{\includegraphics[width=0.45\textwidth]{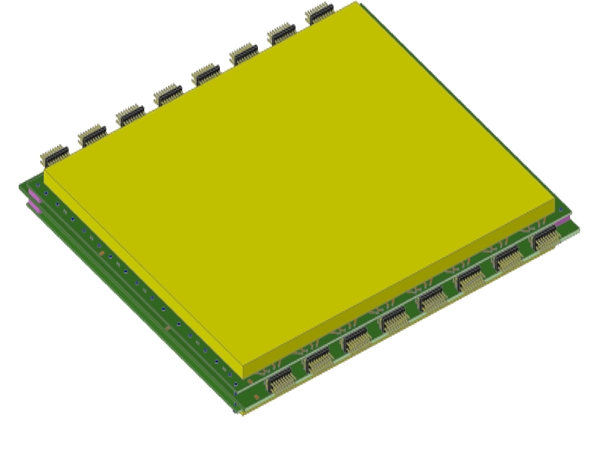}} \quad
	\subfloat[]{\includegraphics[width=0.45\textwidth]{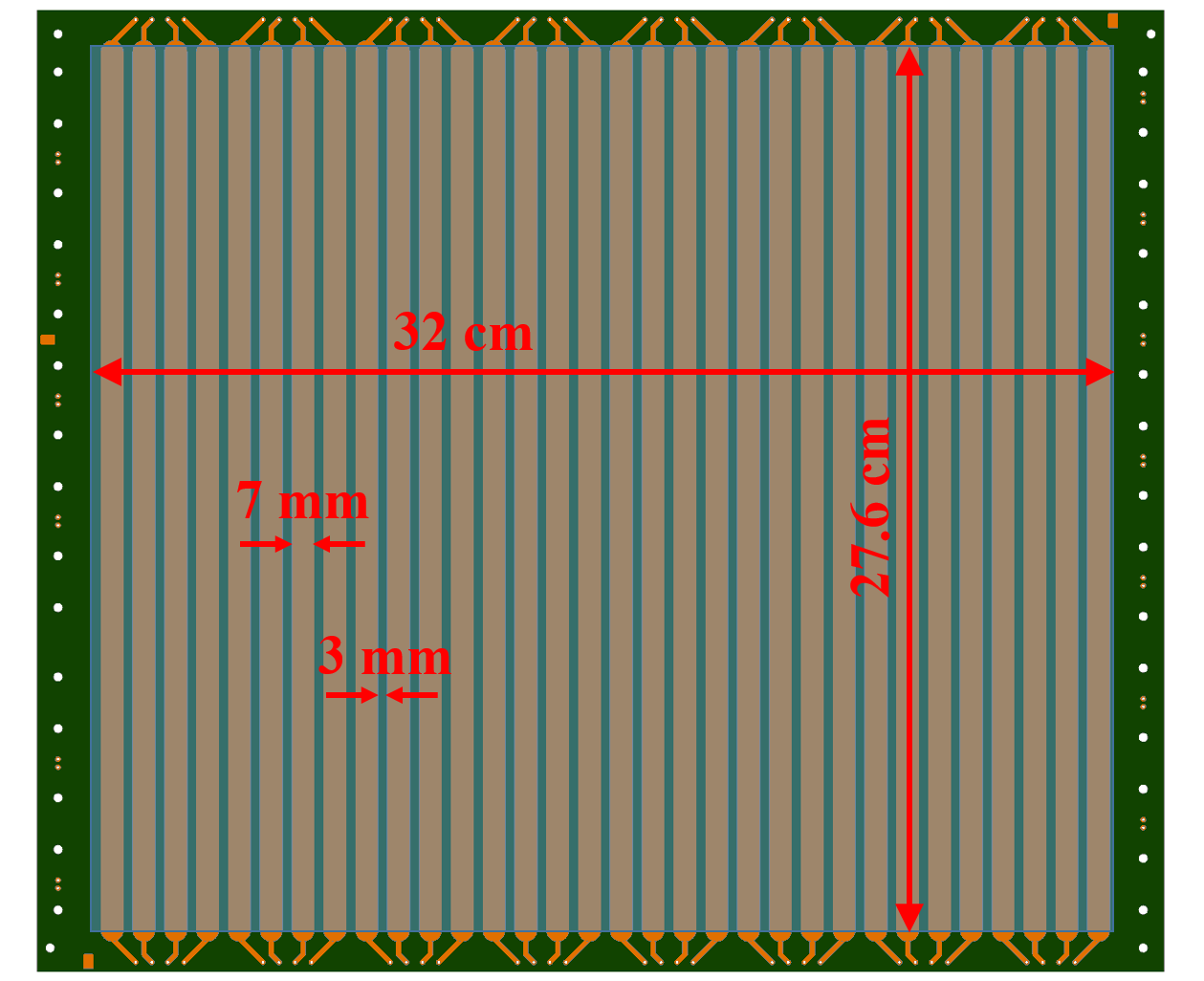}}
	\caption{(a) The schematic of MRPC3b structure. (b) The readout strip  pattern of MRPC3b .}
	\label{fig:2}
\end{figure}

MRPC3b is positioned at the top and bottom of the CBM TOF wall, corresponding to the low rate region as shadowed in Figure \ref{fig1}(b). A system time resolution of below 80 ps and an efficiency of above 95\% are required for the TOF system of CBM \cite{deppner2012cbm}.
Figure \ref{fig:2} shows the schematic of the MRPC3b structure. It is a two-stacks, 10 gas gaps MRPC counter. The resistive plates are made of ultra-thin float glass, merely 0.28 mm thick, to enhance the rate capability \cite{thinfloat}. In each stack, 6 glass sheets are separated by fishing lines to form the gas gaps of 0.23 $\rm mm$. The outer surfaces of the glass stacks are sprayed with graphite layers, serving as the High Voltage (HV) electrodes. Figure \ref{fig:2}(b) shows the readout strip pattern of MRPC3b. It has 32 double-end readout strips covering the active area of 32 $\rm cm$ $\rm \times$ 27.6 $\rm cm$. The pitch of the strips is 1 $\rm cm$ while the strip width is 0.7 $\rm cm$. 
Three layers of readout strips collect the induced signals from the avalanches with the negative signals on the middle layer and the positive ones on the outer layers. The differential signals are then read out by PADI Front End Electronics (FEE) \cite{padi}
boards from both strip ends. In order to suppress the signal reflection, the characteristic impedance of the strips is carefully calculated and designed to be 50 $\rm \Omega$ differential which matches the input impedance of PADI. 

The MRPC3b prototypes were tested at the Beijing Electron-Positron Collider (BEPC) E3 line in 2016 with a 700 MeV hadron 
beam \cite{mrpc3bbeam}. The results show MRPC3b has good performance of time resolution better than 60 ps and efficiency higher than 98\%.

\section{Batch test of 81 MRPC3bs} \label{batch}
\subsection{STAR-eTOF upgrade} \label{upgrade}

\begin{figure}[htbp]
	\centering 

        \includegraphics[width=0.45\textwidth]{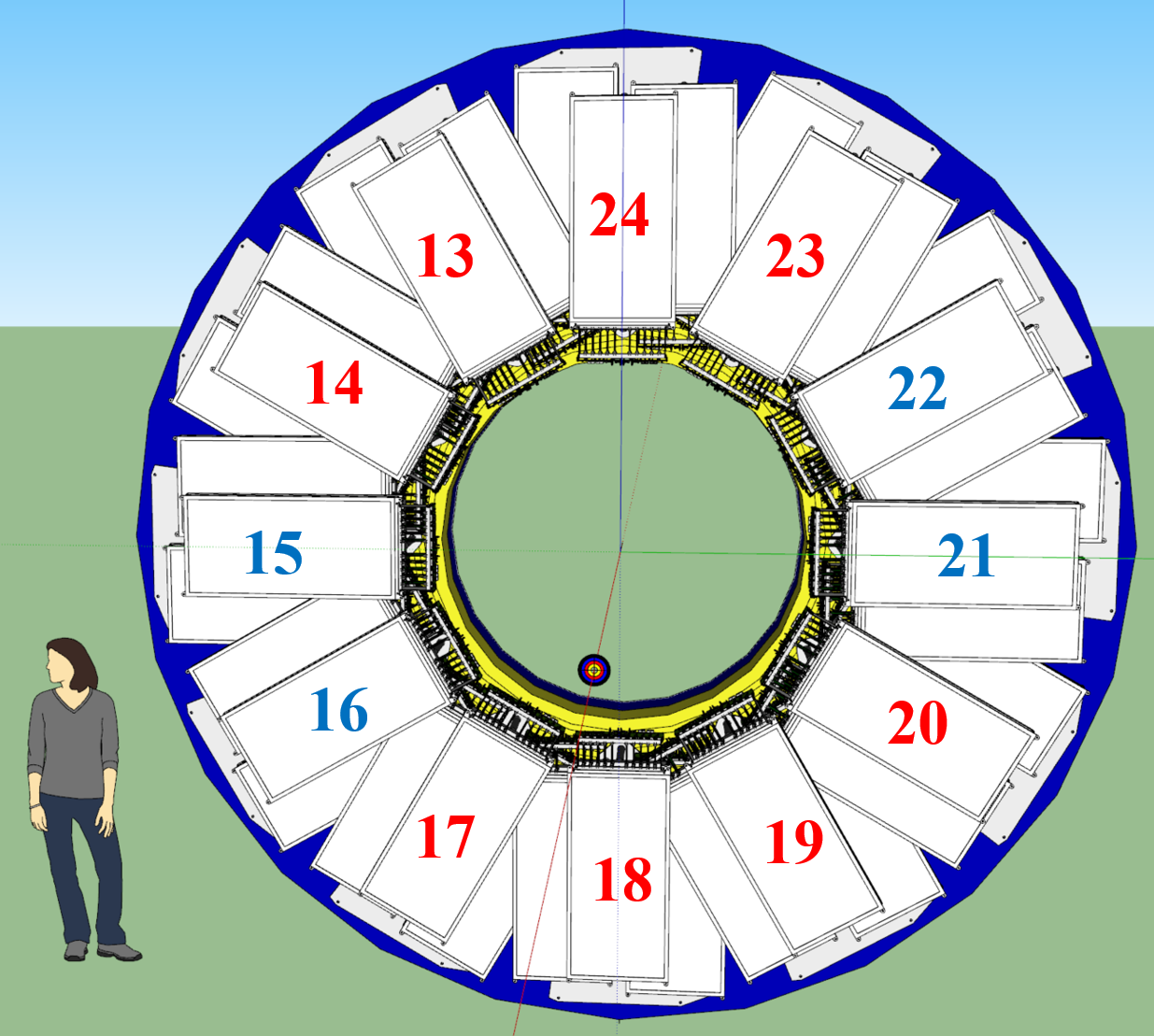}
	\caption{ The conceptual design of eTOF wheel for STAR, numbered corresponding to the STAR inner-TPC sectors.}
	\label{fig:3}
\end{figure}
\begin{figure*}[htbp]
	\centering 
	
	\subfloat[]{\includegraphics[width=0.45\textwidth]{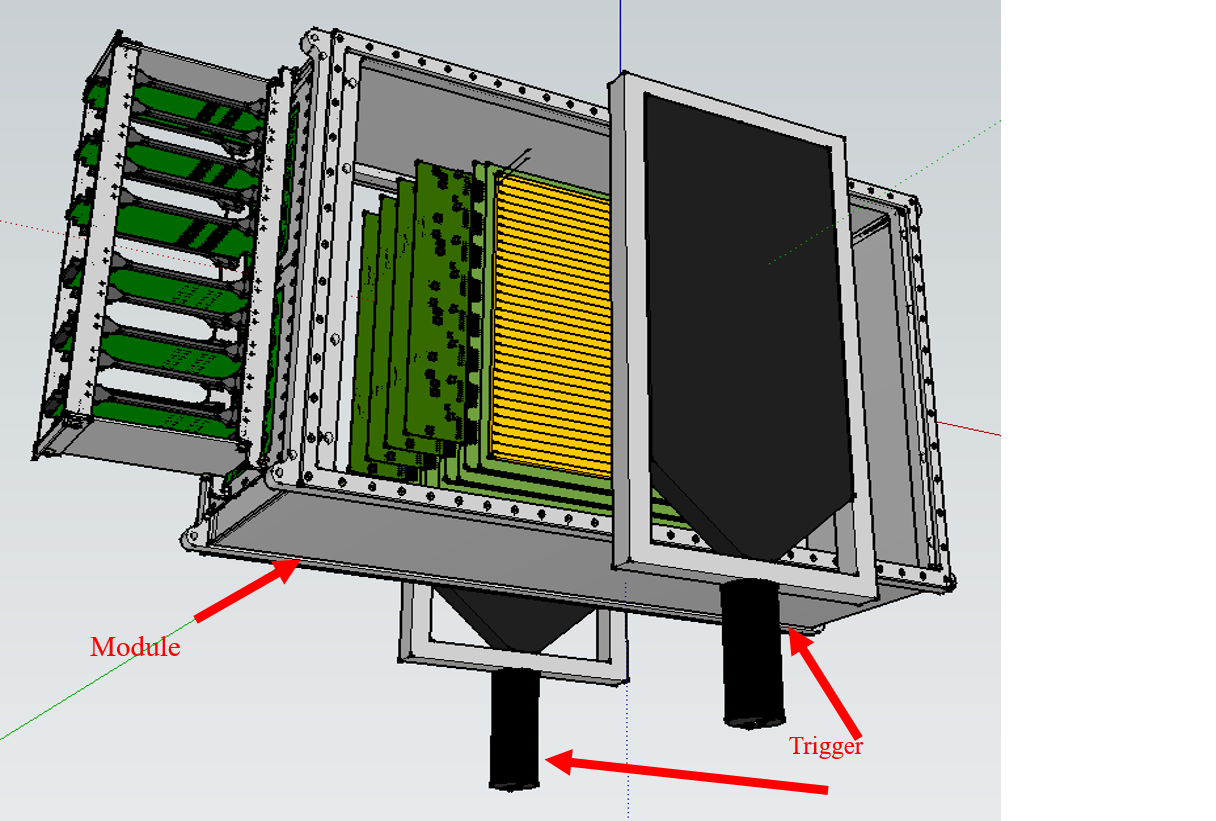}}
	\subfloat[]{\includegraphics[width=0.45\textwidth]{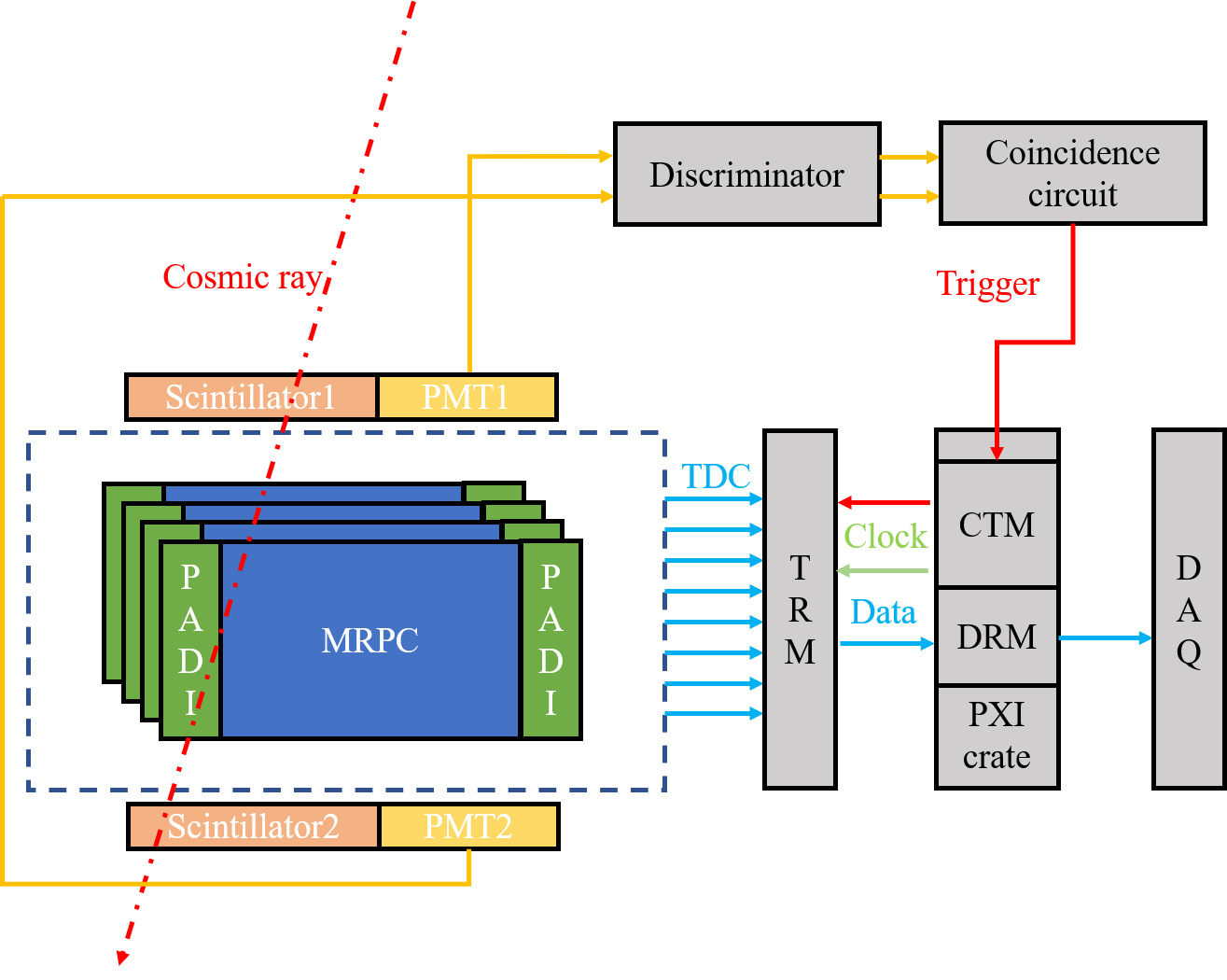}}
	
	\caption{\label{fig:4} The cosmic ray test system for MRPC3b batch test. }
\end{figure*}

The Solenoidal Tracker at RHIC (STAR) experiment at Brookhaven National Laboratory  is an important heavy ion collision experiment in the world. The STAR Collaboration proposed the Beam Energy Scan phase \uppercase\expandafter{\romannumeral2} (BES-II) program with relevant improvements to the STAR detectors \cite{STARBES}. The newly established eTOF is one of the major upgrades \cite{staretof}.
The eTOF system expands the pseudorapidity coverage and provides particle identification (PID) capabilities in the range of -1.6 $\textless \eta \textless$-1.1 for collider collision mode and at mid-rapidity with center-of-mass energies from 3.0 to 4.5 $\rm GeV$ for the fixed target mode.  As part of the FAIR Phase-0 program for CBM-TOF, the STAR-eTOF project provides a unique opportunity to test the counter stability and the commissioning of the CBM-TOF modules before the operation of CBM experiment.
Figure \ref{fig:3} shows the conceptual design of the eTOF wheel for STAR. 
The STAR-eTOF is composed of 12 sectors, each containing 3 modules with 3 MRPC detectors per module. Sectors labeled in blue are comprised of MRPC3as \cite{MRPC3a_mass}, while sectors labeled in red consist of MRPC3bs. In total, 81 MRPC3bs has been constructed and tested in USTC and subsequently sent to Heidelberg for module assembly.
The eTOF wheel had been fully installed at the STAR experiment by November 2018.
\subsection{Batch test platform for MRPC3b} \label{platform}
A cosmic ray test system is built for MRPC3b batch test. 
The system consists of two plastic scintillator counters each coupled 
with one Photomultiplier Tube (PMT) on single end. The size of the scintillator is 20 $\rm cm$ $\rm \times$ 40 $\rm cm$. An aluminum gas tight box, which can accommodate 4 MRPCs orderly, is positioned between the scintillators, as shown in Figure \ref{fig:4}(a). The signals from the two PMTs are initially discriminated by the Low Threshold Discriminator (CAEN N845) and then sent to a coincidence unit to generate a trigger for the system. Two PADI boards are plugged on each end of the MRPC directly to amplify and discriminate the MRPC signals. The back-end signals are processed and acquired by a 320-channel time digitizing  and readout electronic system (shown in Figure \ref{fig:4}(b)) designed by the fast electronics  laboratory of USTC\cite{lichao}. In this system, the discriminated MRPC signals are digitized  by the Time-to-Digital Converter (TDC). Then, the digitized time data are aggregated at the TDC Readout Motherboard (TRM) and transmitted to the Data Readout Modules  (DRM) via optical links. The Clock Trigger Module (CTM)  distributes the clock and trigger to TRM reversely. Finally, the DRMs relay the data to the Data Acquisition System (DAQ) through the  Gigabit Ethernet ports in parallel.   The laboratory test results indicate that the electronics can achieve a time resolution better than 20 ps.

\subsection{MRPC3b Batch test results} \label{results}

81 MRPC3bs have been constructed in USTC with a strict quality assurance (QA) and quality control (QC) process, including materials checking, process checking, and performance checking\cite{massprod}. All these MRPC3bs had been tested in laboratory with the cosmic ray test platform described in \ref{platform} before they were installed at STAR-eTOF.

\begin{figure}[htbp]
	\centering 
	
	\includegraphics[width=0.6\textwidth]{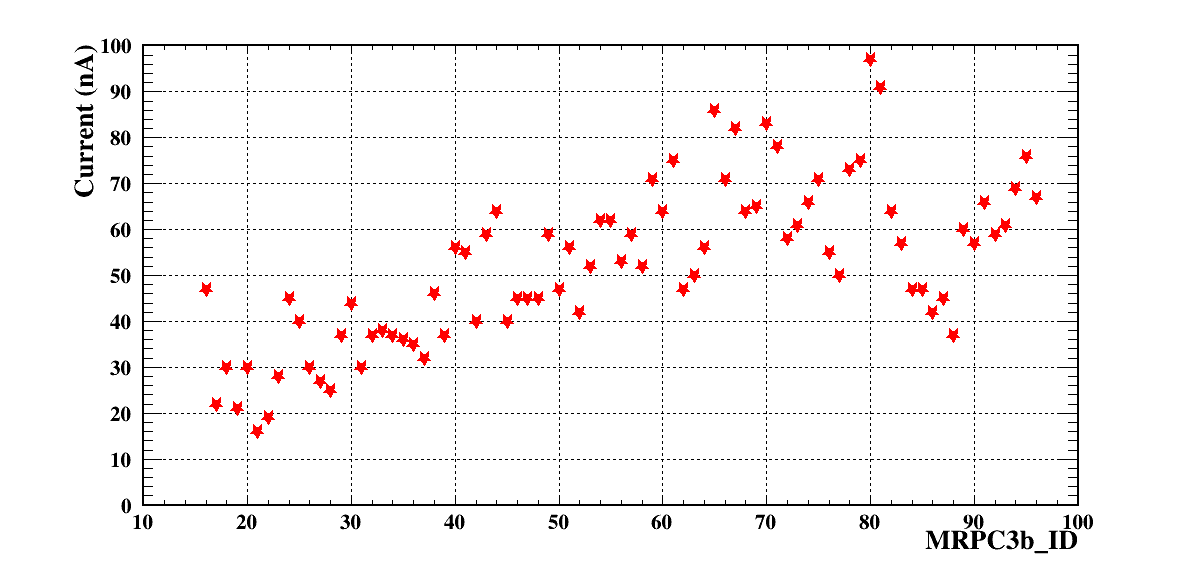}
	
	\caption{ Working current under HV of $\rm \pm$6000 V of all the MRPC3bs.}
	\label{fig:5}
\end{figure}

For the MRPC3b performance check, we test the current of all the counters under the working condition. The working gas mixture is composed of 95\% Freon and 5\% $\rm iso$-$\rm C_4H_{10}$. According to the former  beam test and cosmic test results\cite{mrpc3b,mrpc3bbeam}, the operating HV is set  to $\rm \pm$6000 $\rm V$ and the PADI threshold is set to -347.2 $\rm mV$. Figure \ref{fig:5} shows statistical working current under HV of $\rm \pm$6000 $\rm V$ for all the 81 MRPC3bs, whose serial numbers ranged from $\#$16 to $\#$96.  The current is read out by WIENER HV module (EHS8280P/N) with 1 $\rm nA$ accuracy.  The results demonstrate that the currents for all MRPC3bs remained below 100 $\rm nA$,  but they increase with MRPC3b serial number.  This trend might be caused by the variations in humidity during the batch test phase. The batch testing process spanned approximately six months, from spring to summer, during which humidity increased accordingly in Hefei, China.

\begin{figure}[htbp]
	\centering 
	
	\subfloat[]{\includegraphics[width=0.35\textwidth]{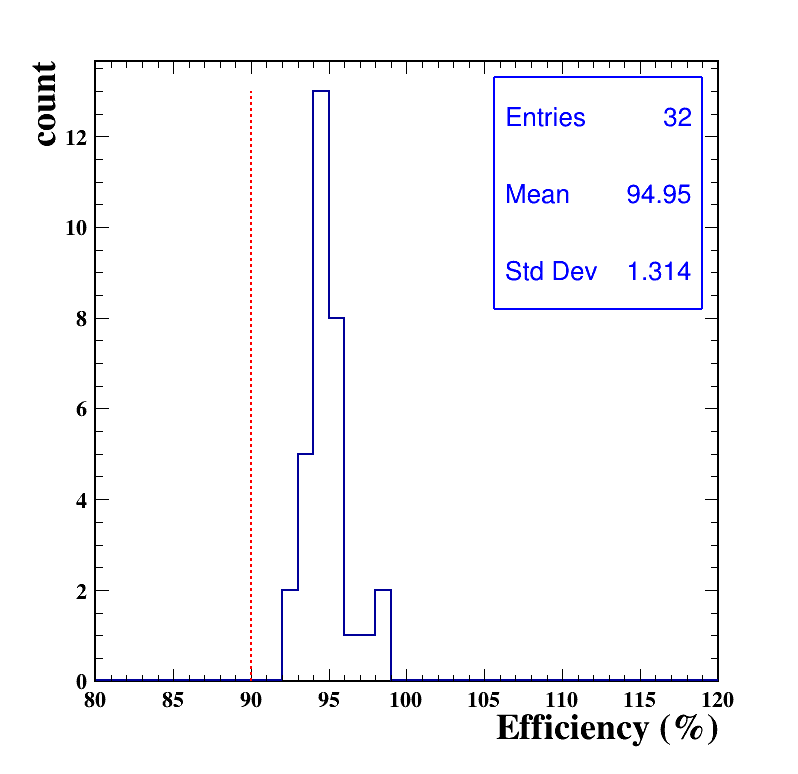}}\quad
	\subfloat[]{\includegraphics[width=0.35\textwidth]{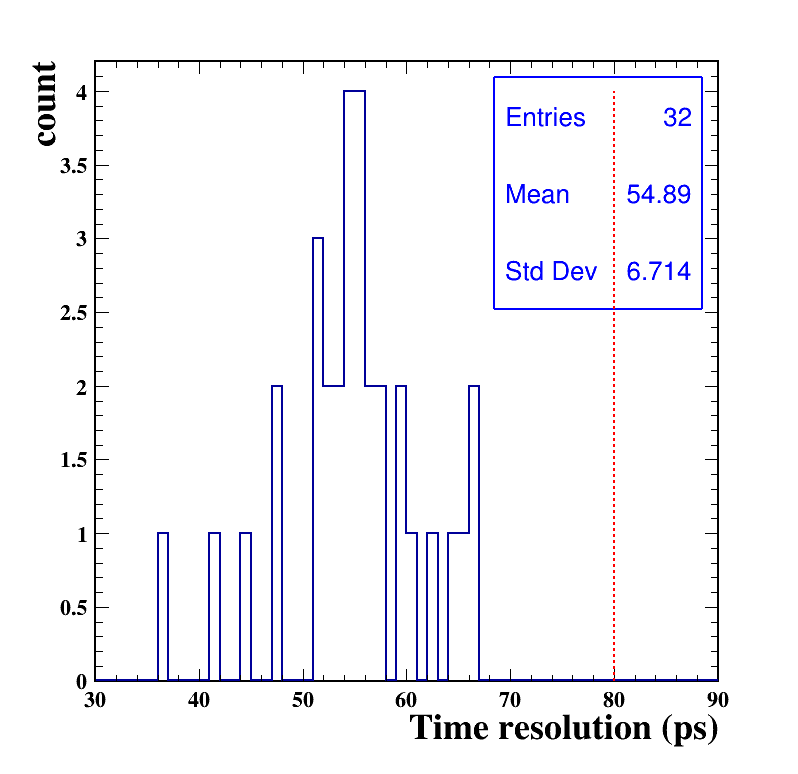}}

	\caption{\label{fig:6} (a) Efficiency of 32 tested MRPC3bs, with the red dashed line representing 90\% efficiency. (b) Time resolution of the 32 tested MRPC3bs, with the red dashed line representing a time resolution of 80 ps. }
\end{figure}

Due to the time constraint for the counter mass production, only a subset of the MRPC3bs could be tested for time resolution and efficiency performance. We randomly selected and tested 32 MRPC3bs out of the total 81 counters. The statistical results of efficiency (Figure \ref{fig:6}(a)) and time resolution (Figure \ref{fig:6}(b)) are shown. From the plots we can see that the mean value of efficiency is around 95\% and the time resolution is better than 70 ps with a mean value of $\sim$ 55 ps. These performances meet the CBM-TOF requirements.

\begin{figure}[htbp]
	\centering 
	
	\subfloat[]{\includegraphics[width=0.35\textwidth]{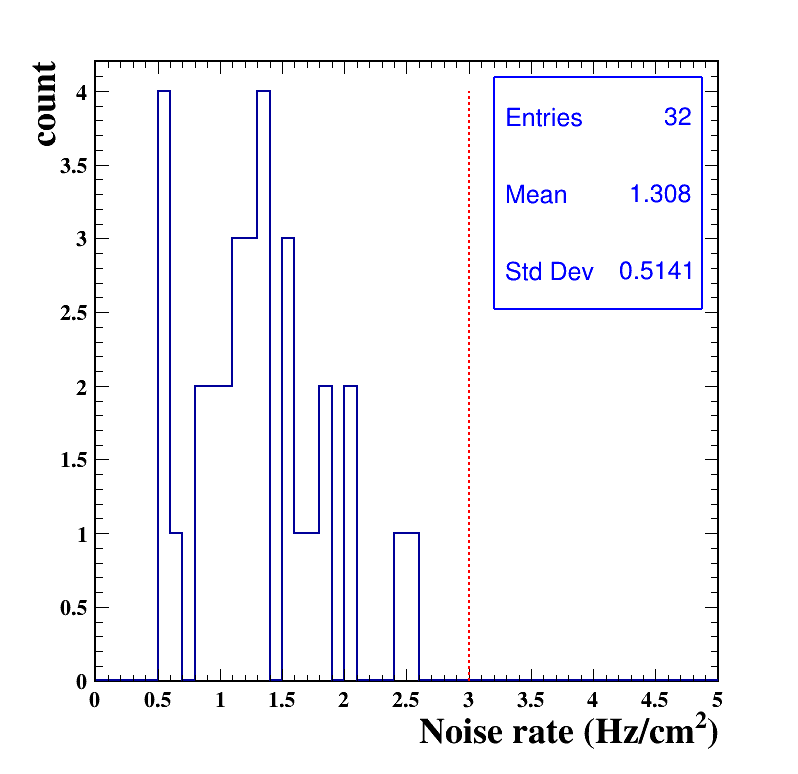}}
	\quad
	\subfloat[]{\includegraphics[width=0.48\textwidth]{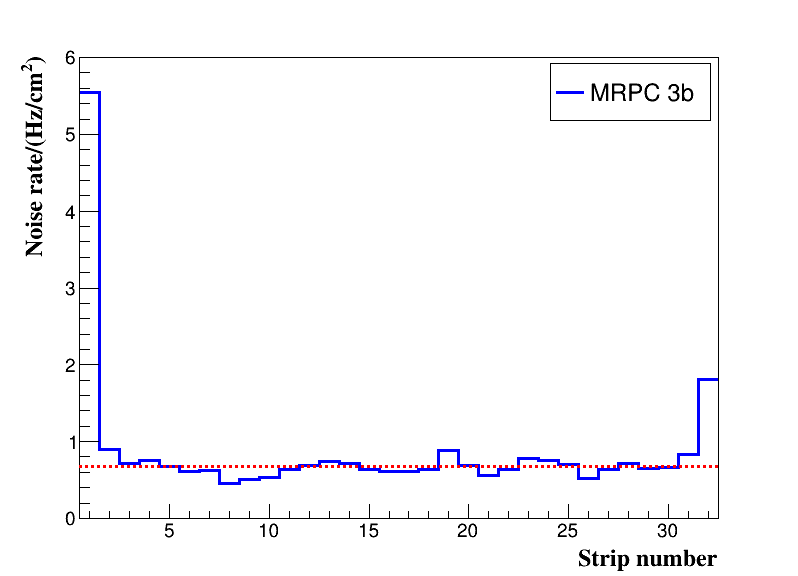}}
	
	\caption{\label{fig:7} (a) Statistics of the noise rate of 32 MRPC3bs, with the red dashed line representing a   3.0 $\rm Hz/cm^2$ noise rate. (b) A Typical noise rate distribution as function of strip serial number,  strip  number is defined from the HV injection side in ascending numerical order, with the red dot line representing the average value of noise rate except for 2 end strips. }
\end{figure}

The noise rate is an important parameter of MRPC, especially for the free streaming readout mode of the CBM experiment. 
To evaluate the noise rate, we utilize the batch test system with a random trigger generated by the CAEN V1718 module\cite{V1718}. 
During the testing process, if the DAQ system records an event with signals detected at both ends of a strip within the matching window of the random trigger signal, it is classified as a noise count on that specific strip. The corresponding noise rate is evaluated in this way.  Figure \ref{fig:7}(a) shows statistical analysis of the noise rate of the 32 MRPC3bs. All counters show a very low noise rate, with an average value as low as 1.3 $\rm Hz/cm^2$ across the entire active area.

In general, all the tested MRPC3bs pass the quality inspections and show excellent performance. But it is noticed that the noise rates of the two edge strips are significantly higher than the others. Specifically, the near-end strip, which is the strip closest to the HV injection side, exhibits a noise rate approximately 8 times higher, while the far-end strip, which is the strip farthest from the HV injection side, shows a noise rate approximately 2 times higher. This is illustrated in Figure \ref{fig:7}(b), where the strip serial number is defined in ascending numerical order from the HV injection side. For a better understanding of this effect, further investigations have been carried out, which will be discussed in the next section.

\section{Noise rate study} \label{noise}
\begin{figure}[htbp]
	\centering 
	
	\includegraphics[width=0.4\textwidth]{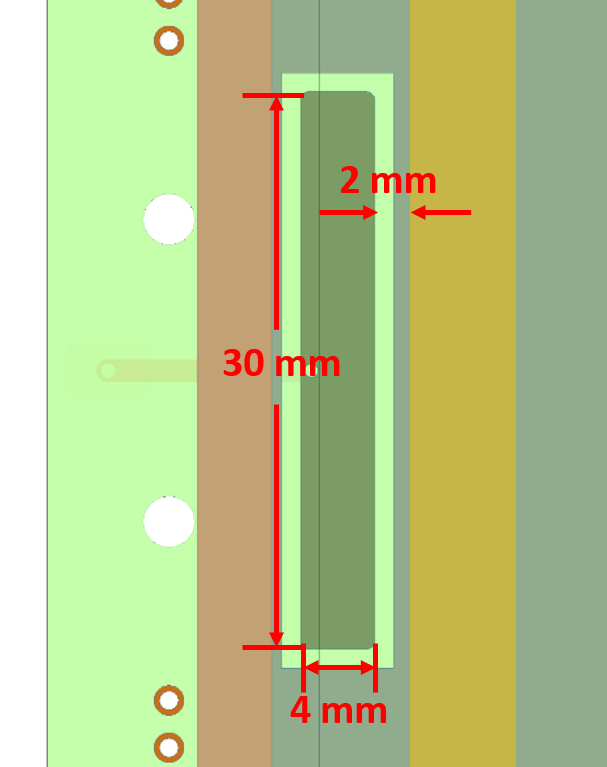}

	\caption{\label{fig:8} The design of the HV injection pad.}
\end{figure}

From the batch test, we found that the two edge strips show much higher noise rates than the others. This phenomenon is suspected to be caused by the cross-talk from the HV source and the edge effect. Figure \ref{fig:8} shows the design details around the HV injection point -- a copper pad with a size of 4 mm $\times$ 30 mm. The distance between the HV pad and the closest strip is merely 2 mm. Simulations are carried out to study the noise rate issue correlated with design parameters and prototype MRPCs are built and tested following the simulation results.

\subsection{Simulation settings and results} \label{sim}

\begin{figure}[htbp]
	\centering 
	
	\includegraphics[width=0.6\textwidth]{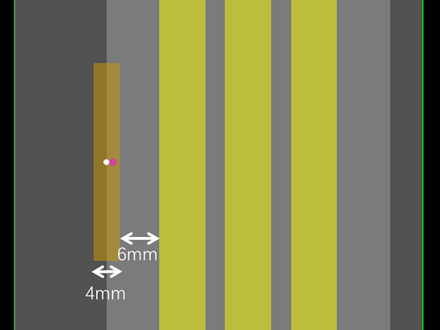}

	\caption{ A typical simplified simulation model of MRPC3b focusing on the HV pad.}
	\label{fig:9}
\end{figure}

In order to study the cross-talk related to the HV pads, simulations 
are carried out using the PCB Studio of CST Suite. Figure \ref{fig:9} shows 
a typical simplified PCB model built in the simulation environment. The PCB model used in the simulation consists of three layers: copper signal strips, copper HV connecting pad, and a graphite layer, arranged from top to bottom. Each pair of adjacent layers is separated by a corresponding insulating layer or masking layer, with the thicknesses employed in the simulation matching those of the actual MRPC3b. Since we focus mainly on the HV pad and the nearby strips, the corresponding models apply only 3 strips with a 1 $\rm cm$ pitch and a 3 mm interval in-between strips. 
The length of the strips is set to 27.6 $\rm cm$, identical to the design parameters of the MRPC3b.

\begin{figure}[htbp]
	\centering 
	
	\subfloat[]{\includegraphics[width=0.25\textwidth]{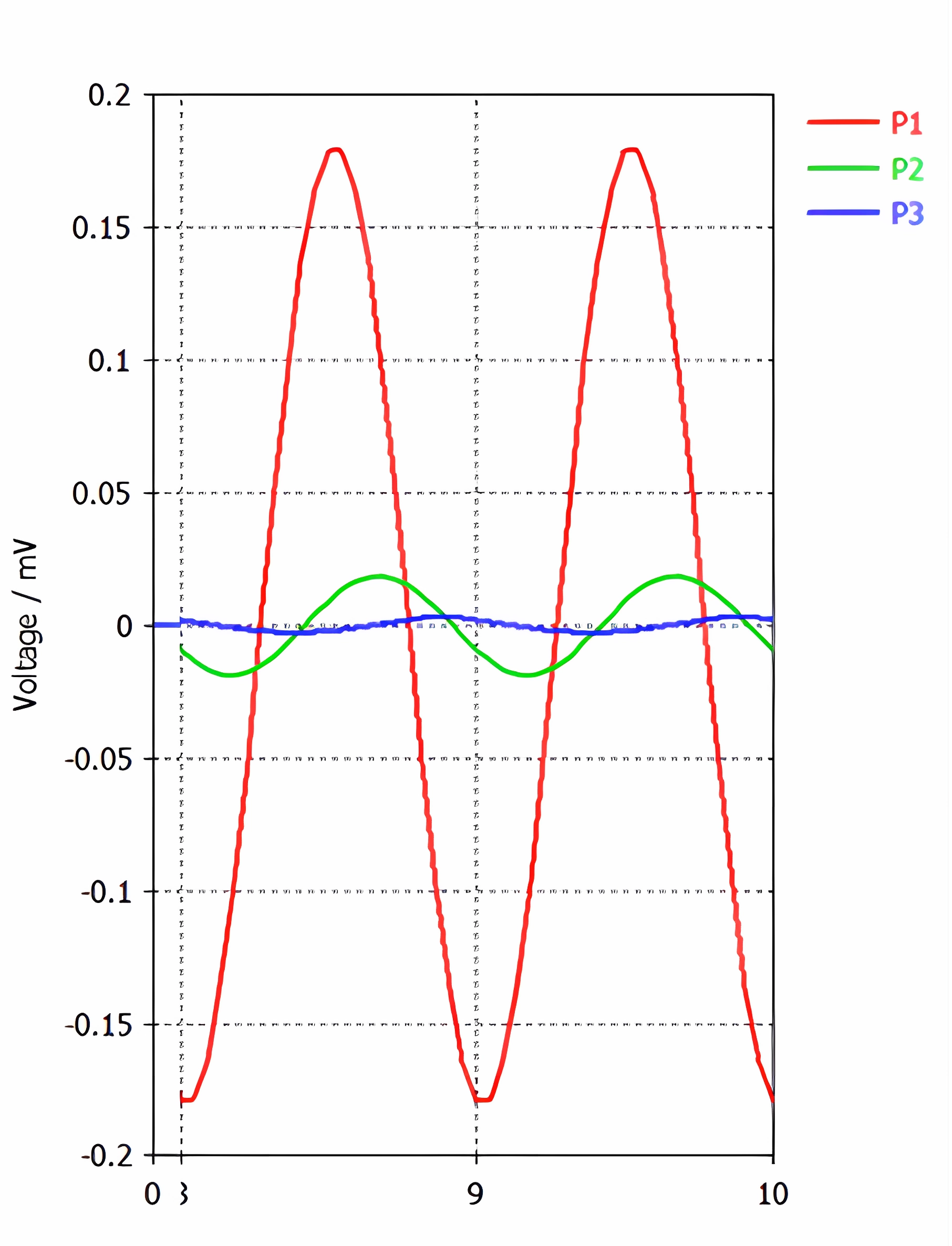}}\qquad
	\subfloat[]{\includegraphics[width=0.25\textwidth]{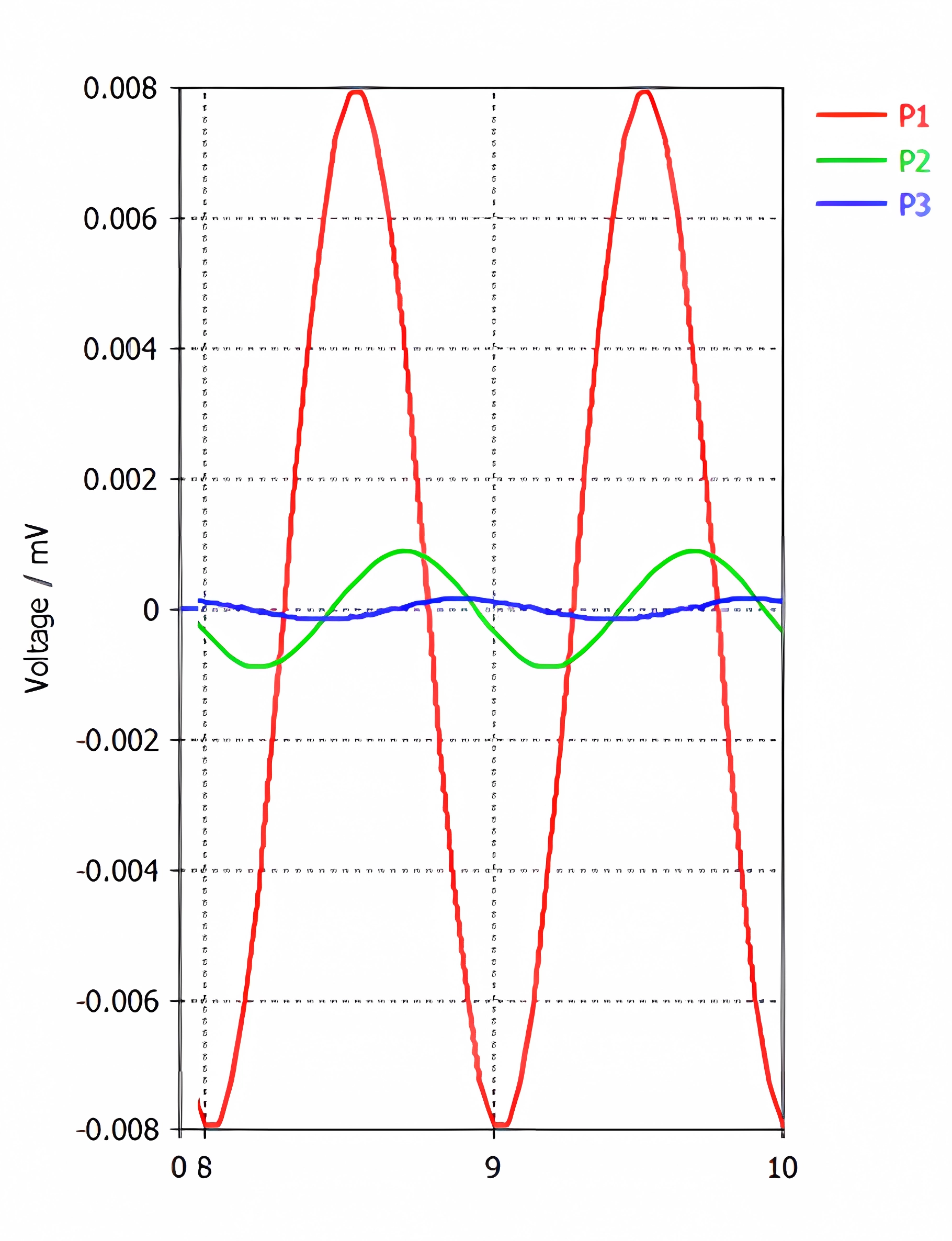}}\quad
	\subfloat[]{\includegraphics[width=0.34\textwidth]{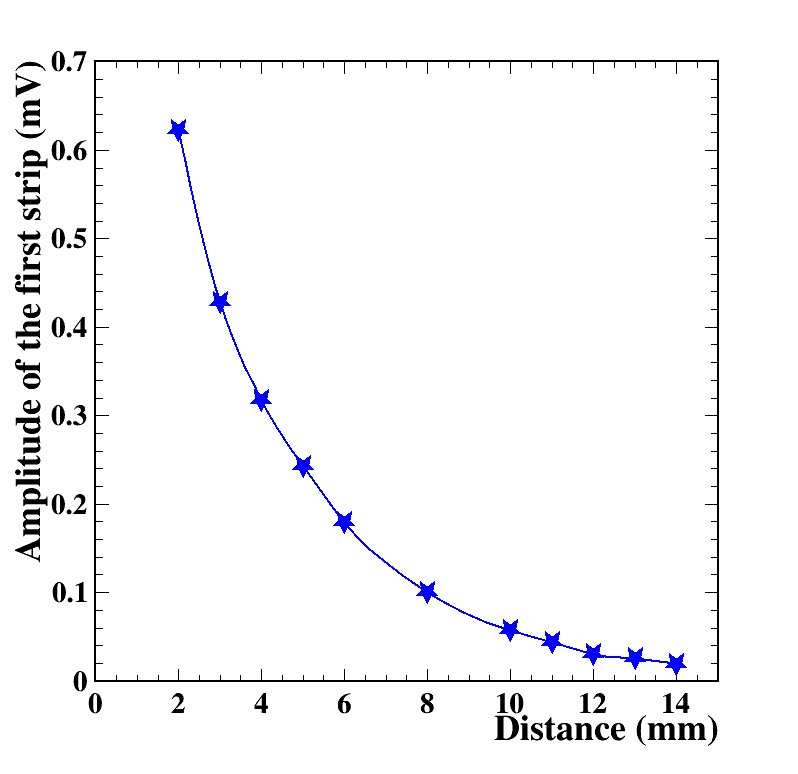}}
	
	\caption{ (a) Simulation results of output signals for a model with 6 mm distance between the HV pad and the strips. 
		(b) Output signals of three strips with stimulus source injected directly into the graphite layer at a 6 mm distance to the strips. 
		(c) Output signal amplitudes of the nearest strip as a function of the distance to the HV pad.}
	\label{fig:10}  
\end{figure}

In this simulation, a 5 mV sinusoidal modulated voltage source is injected 
into the HV pad and the HV pad is connected to the graphite layer.
Firstly, the HV pad is set to a 4 mm $\times$ 30 mm rectangular shape as the 
real MRPC3b. The distance between the HV pad and the nearest strip varies from 2 mm to 14 mm.
Figure \ref{fig:10}(a) shows typical output signals of three readout strips 
when the distance is 6 mm. The red line P1 is the signal read out from the nearest 
strip and the amplitude is 0.17 mV. The green and blue lines show the signals of 
the second and third strips respectively, whose amplitudes are much lower. 
As a comparison, if the same stimulus source is injected into the graphite 
layer directly at the same distance without the HV pad, the output signals 
are shown in Figure \ref{fig:10}(b). The signal amplitude from the first strip 
is about 0.008 mV, which is much smaller than the model with an HV pad. 
These results indicate the HV pad has a great impact on the nearest strip but little on the others. 
Figure \ref{fig:10}(c) summarizes the signal amplitudes of the nearest strip 
as a function of the distance to the HV pad. The amplitudes of the signals decrease as the distance increase. 

Several other models are also built in the simulation by changing the length, size, and shape of the HV pad. 
Table \ref{tab:i} lists the detailed structure information and simulated amplitudes on the nearest strip of the four models. 
The distances between the HV pad and the strips are fixed at 6 mm for all these models. 
The HV pads of R6-1 and S6 have the same area, but the length of S6 is half of R6-1. 
The HV pad of R6-2 has the same length as S6 and the same width as R6-1. 
For the cases of R6-2 and S6, whose length is the same, the results are very close to each other. But for the longer HV pad, R6-1, the amplitude is a bit larger.
These results hint that the length of the HV pad affects the cross-talk relevantly. 
The circular case with a radius of 6 mm, C6, shows the smallest cross-talk effect of 0.08 mV amplitude.

\begin{table*}[htbp]
	\centering
	\caption{\label{tab:i} Simulation results of different models.}
	\smallskip
 \resizebox{1\columnwidth}{!}{
	\begin{tabular}{ccccc}
		\hline
		Model & R6-1 & R6-2 & S6 & C6 \\ \hline
		HV pad & Rectangle & Rectangle & Rectangle & Circle \\ 
		(size) & (4 mm$\times$30 mm) & (4 mm$\times$15 mm) & (8 mm$\times$15 mm) & (r = 6 mm) \\ 
		Amplitude (mV) & 0.17 & 0.11 & 0.10 & 0.08 \\ \hline
	\end{tabular}
 }
\end{table*}

From  the simulation, it is clearly indicated that the distance between HV pad and strips 
and the dimension of the HV pad impact the cross-talk on the nearest strip.
Noise rate tests on different prototypes are carried out to demonstrate the simulated results.

\subsection{Prototype test} \label{proto}
\begin{figure}[htbp]
	\centering 
	
	\includegraphics[width=0.60\textwidth]{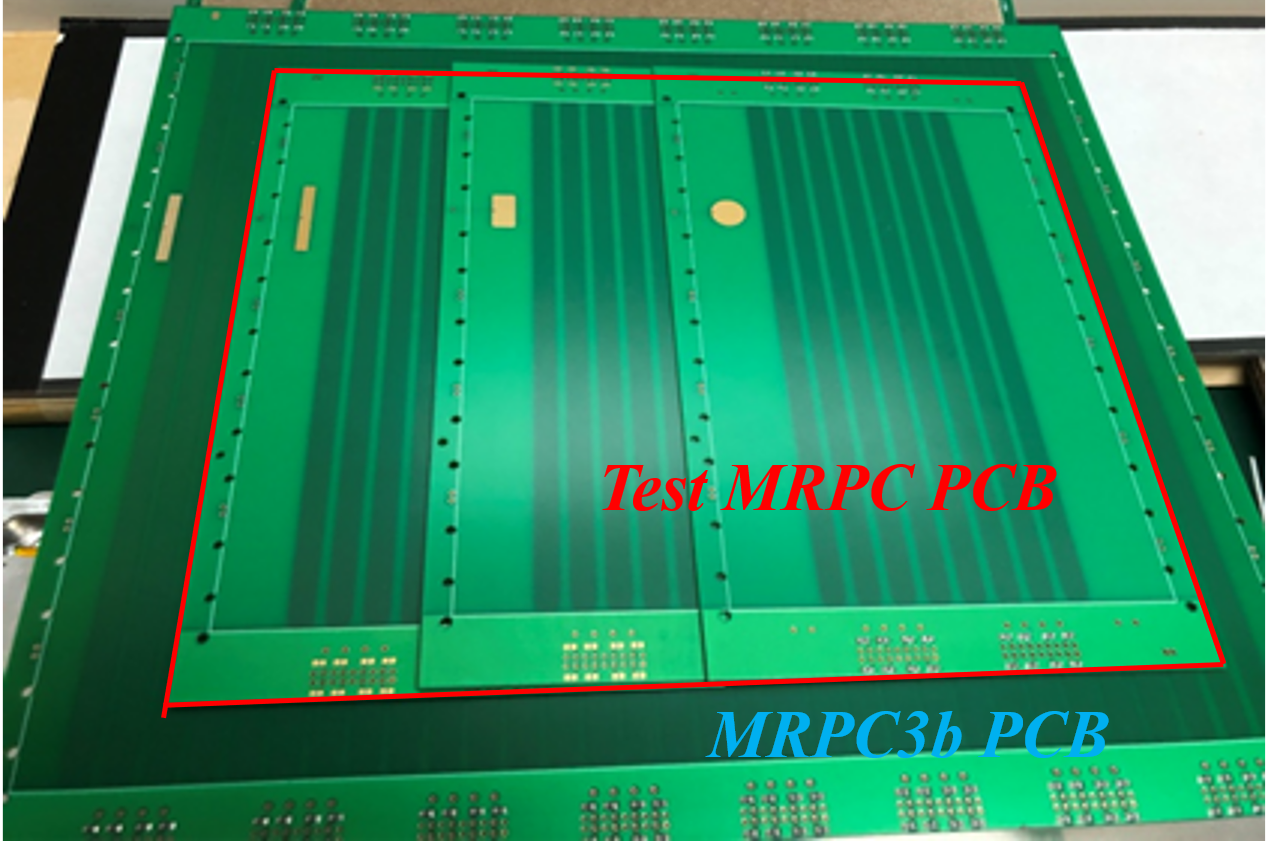}

	\caption{ Photo of the prototype PCBs on top of the MRPC3b PCB.}
	\label{fig:11}
\end{figure}

Several MRPC prototypes are designed and built according to the simulation results.
The detailed structural information of these prototypes is provided in Table \ref{tab:ii}, allowing for a convenient comparison of the effects resulting from variations in the distance between the HV pad and the nearest strip, as well as the shape of the HV pad. 
Figure \ref{fig:11} shows the photo of different types of designed PCBs.  
All the prototypes have 8 strips, each measuring 20 cm in length, with a pitch of 1 cm and an interval of 0.3 cm. 
To examine the influence of the edge effect, the distance between the glass edge and the far-end strip has been increased, ranging from 2.1 cm to 3.1 cm. This is considerably larger than the distance used in the previous design of MRPC3b, which was 0.7 cm. 
They closely resemble MRPC3b in terms of structure, with the only difference being the size.

Among the prototypes, R4, R6-1, and R8 have varying distances between the HV pad and the nearest strip, while maintaining the same HV pad shape. Conversely, R6-1, S6, and C6 possess the same distance but exhibit different HV pad shapes. Subsequently, noise rate tests are conducted on all the constructed MRPC prototypes using the system discussed in Section \ref{platform}. 

\begin{table*}[!htbp]
	\centering
	\caption{\label{tab:ii} Structural information of PCBs.}
	\smallskip
 \resizebox{1\columnwidth}{!}{
	\begin{tabular}{ccccccc}
		\hline
		Model\centering  & R4 & R6-1 & R8 & S6 & C6 \\ \hline
		\multirow{2}{*}{HV pad}\centering & Rectangle & Rectangle & Rectangle & Rectangle & Circle \\ 
		~ & (4$\times$30 mm) & (4$\times$30 mm) & (4$\times$30 mm) & (8$\times$15 mm) & (r=6 mm) \\ 
		\makecell{Distance  between HV pad \\and near-end strip }\centering  & 4 mm & 6 mm & 8 mm& 6 mm & 6 mm \\ 
		\makecell[c]{Distance  between glass edge \\and far-end strip} & 31 mm & 29 mm & 27 mm & 25 mm & 21 mm \\ \hline
	\end{tabular}
 }
\end{table*}

Table \ref{tab:iii} lists the test results of all the prototypes. It can be observed that the noise rates of the far-end strips in all the tested prototypes are very low and comparable to the average of the other strips. This phenomenon is likely due to the reduced edge effect achieved in the design. 
For the near-end strips, a comparison of the MRPC3b, R4, R6-1, R8 prototypes reveals a significant decrease in the noise rate as the distance between the HV pad and the strip increases. This observation is consistent with the simulation results. The noise rate of R6-1 is slightly lower than that of R8, which could be due to the amplitude change becomes negligible when the distance exceeds 6 mm, as indicated by the previous simulation. Comparing R6-1, S6, and C6, we find that the circular pad C6 prototype has the smallest noise rate. However, since the average noise rate level differs across detectors and strips, the impact of these prototypes on HV cross-talk appears to be negligible. Overall, these test results are well consistent with the findings from the previous simulations.
\begin{table*}[htbp]
	\centering
	\caption{\label{tab:iii} The noise rate test results of different prototypes.}
	\smallskip
 \resizebox{1\columnwidth}{!}{
	\begin{tabular}{cccccccc}
		\hline
		Model &  & MRPC3b & R4 & R6-1 & R8 & S6 & C6 \\ \hline
		\multirow{3}{*}{Noise Rate( Hz/$\rm cm^2$)}  & Far-end strip & 1.80 & 0.58 & 0.52 & 0.57 & 0.42 &0.31 \\ 
		~ & Near-end strip & 5.54 & 1.09 & 0.36 & 0.59 & 0.49 &0.29\\
		~ & Average of other strips & 0.67 & 0.68 & 0.47 & 0.57 & 0.31 &0.39\\ \hline
	\end{tabular}
 }
\end{table*}

\subsection{Optimization for MRPC3b} \label{opt}

\begin{figure}[htbp]
	\centering 
	
	\subfloat[]{\includegraphics[width=0.27\textwidth]{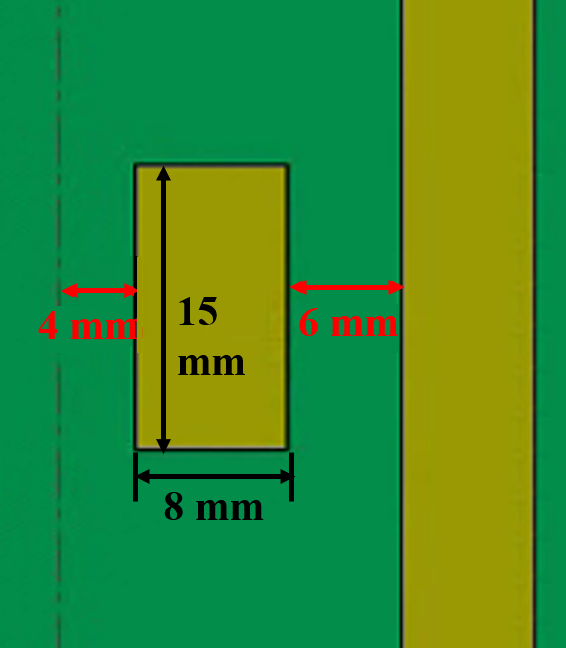}}\quad
	\subfloat[]{\includegraphics[width=0.45\textwidth]{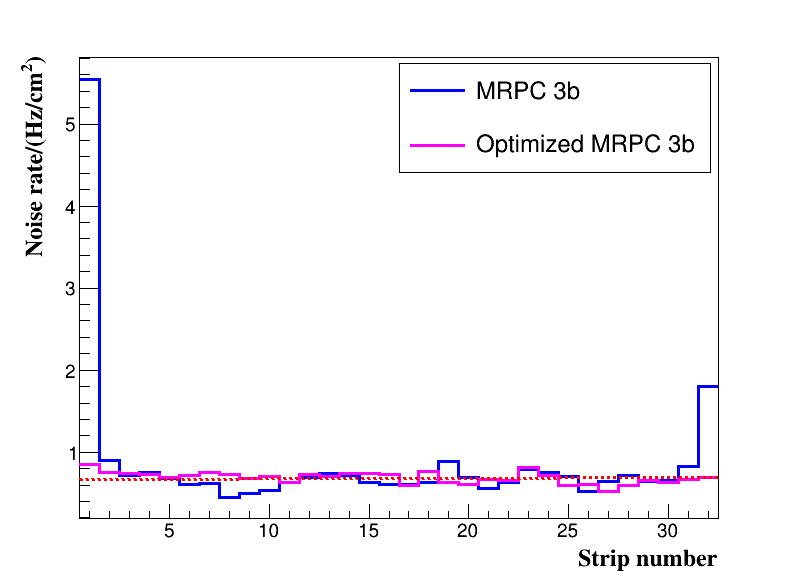}}
	
	\caption{(a) The optimized design of MRPC3b's PCB structure.
		(b) The noise rate comparison of the optimized and the original MRPC3b, with the dot line representing the average value of optimized MRPC3b noise rate except for 2 end strips.}
	\label{fig:12}  
\end{figure}

To reduce the noise rate on edge strips, the structure of MRPC3b has been optimized based on the simulation and test results. The optimized structure is depicted in Figure \ref{fig:12}(a). This optimization takes into account the edge effect, cross-talk of the HV pad, and the requirements for the active area. In the optimized design, the HV pad is rectangular in shape, measuring 8 mm $\times$ 15 mm. The distance between the HV pad and the first strip is set to 6 mm, while the distance from the glass edge to both end strips is increased to 18 mm. To accommodate this design, the size of the glass has also been increased accordingly. It is important to note that this improvement has little impact on the effective coverage of the CBM-TOF module. The MRPC detectors in the CBM-TOF module are rotated towards the target point, resulting in the overlapping regions that cover the inefficient area. 

The optimized MRPC3b has been assembled and tested to evaluate the noise rate. Figure \ref{fig:12}(b) provides a comparison between the optimized MRPC3b and the original MRPC3b, clearly demonstrating a significant reduction in the noise rate of the near-end and far-end strips. Specifically, the near-end strip exhibits a noise rate of 0.87 $\rm  Hz/cm^2$, the far-end strip shows a noise rate of 0.69 $\rm  Hz/cm^2$, both of which have significantly decreased compared to the previous design. Additionally, the noise rates of the edge strips are now comparable to the average noise rate of 0.68 $\rm  Hz/cm^2$ for the other strips. This validates the effectiveness of the approach in reducing noise rates at the edge strips. The optimized structure will be implemented in the CBM-TOF wall construction.

\section{Conclusion} \label{con}
In conclusion, the mass production of 81 MRPC3b counters for the STAR-eTOF upgrade has been successfully completed. HV training and batch testing using a cosmic ray test system have demonstrated the overall excellent performance of the counters. The statistical efficiency was found to be approximately 95\%, and all tested counters exhibited a time resolution better than 70 ps, indicating their suitability for the intended application.

During the testing phase, an issue regarding the abnormal increase in noise rate on the edge strips was observed. Through simulation using CST Studio and subsequent prototype tests, the influence of HV pad crosstalk and edge effects on the noise rate was investigated. The optimized MRPC3b structure, derived from the simulation and validated through prototype testing, has effectively addressed this issue. The noise rates of the edge strips have been brought to levels comparable to those of the middle strips, ensuring a more uniform performance across all strips. The optimized MRPC3b sets the stage for its implementation and will be installed in the future CBM-TOF wall.

\section*{Acknowledgments}
The authors thank the high energy physics group of USTC. This project is supported by National Key Programme for S\&T Research and Development under Grant NO. 2018YFE0205202, National Natural Science Foundation of China under Grant No. 11975228 and 12205296, and the State Key Laboratory of Particle Detection and Electronics under Grant No. SKLPDE-ZZ-202320.






\bibliography{reference1}





\end{document}